\documentclass[a4paper,10pt,preprint,twocolumn,3p]{elsarticle}

\usepackage{amsthm}
\usepackage{mathtools}
\usepackage{physics}
\usepackage{graphicx}
\usepackage{adjustbox}
\usepackage{placeins}
\usepackage[T1]{fontenc}
\usepackage{multirow}
\usepackage{graphicx}
\usepackage{dcolumn}
\usepackage{bm}
\usepackage{color}
\usepackage{comment}
\usepackage{booktabs}
\usepackage[dvipsnames]{xcolor}
\usepackage[pdftex,colorlinks,citecolor=blue,urlcolor=blue,linkcolor=blue,bookmarks]{hyperref}

\journal{Physics Letters B}

\biboptions{numbers,sort&compress}
\begin{document}
\title{Two-neutrino $\beta\beta$ decay of $^{136}$Xe to the first excited $0^+$ state in $^{136}$Ba}

\author[add1]{L. Jokiniemi}
\ead{ljokiniemi@triumf.ca}
\author[add2]{B. Romeo}
\ead{bromeo@dipc.org}
\author[add5,add6,add7]{C. Brase}
\ead{cbrase@theorie.ikp.physik.tu-darmstadt.de}
\author[add8,add10,add9]{J. Kotila}
\ead{jenni.kotila@jyu.fi}
\author[add3,add4]{P. Soriano}
\ead{pablosoriano@ub.edu}
\author[add5,add6,add7]{A. Schwenk}
\ead{schwenk@physik.tu-darmstadt.de}
\author[add3,add4]{J. Men\'{e}ndez}
\ead{menendez@fqa.ub.edu}    

\address[add1]{TRIUMF, 4004 Wesbrook Mall, Vancouver, BC V6T 2A3, Canada}
\address[add2]{Donostia International Physics Center, 20018 San Sebasti\'an, Spain}
\address[add5]{Technische Universit\"at Darmstadt, Department of Physics, 64289 Darmstadt, Germany}
\address[add6]{ExtreMe Matter Institute EMMI, GSI Helmholtzzentrum f\"ur Schwerionenforschung GmbH, 64291 Darmstadt, Germany}
\address[add7]{Max-Planck-Institut f\"ur Kernphysik, Saupfercheckweg 1, 69117 Heidelberg, Germany}
\address[add8]{Finnish Institute for Educational Research, University of Jyväskylä, P.O. Box 35, Jyväskylä FI-40014, Finland }
\address[add10]{Department of Physics, University of Jyväskylä, P.O. Box 35, Jyväskylä FI-40014, Finland}
\address[add9]{Center for Theoretical Physics, Sloane Physics Laboratory, Yale University, New Haven, Connecticut 06520-8120, USA}
\address[add3]{Departament de Física Quàntica i Astrofísica, Universitat de Barcelona, 08028 Barcelona, Spain}
\address[add4]{Institut de Ciències del Cosmos, Universitat de Barcelona, 08028 Barcelona, Spain}

\begin{abstract}
We calculate the nuclear matrix element for the two-neutrino $\beta\beta$ decay of $^{136}$Xe into the first excited $0^+$ state of $^{136}$Ba. We use different many-body methods: the quasiparticle random-phase approximation (QRPA) framework, the nuclear shell model, the interacting boson model (IBM-2), and an effective field theory (EFT) for $\beta$ and $\beta\beta$ decays. While the QRPA suggests a decay rate at the edge of current experimental limits, the shell model points to a half-life about two orders of magnitude longer. The predictions of the IBM-2 and the EFT lie in between, and the latter provides systematic uncertainties at leading order. An analysis of the running sum of the nuclear matrix element indicates that subtle cancellations between the contributions of intermediate states can explain the different theoretical predictions. For the EFT, we also present results for two-neutrino $\beta\beta$ decays to the first excited $0^+$ state in other nuclei.
\end{abstract}

\maketitle

\section{Introduction}

Two-neutrino double-beta ($2\nu\beta\beta$) decay and two-neutrino double-electron capture ($2\nu$ECEC) change the atomic number of a nucleus by two via emission or capture of two electrons accompanied by emission of two neutrinos. These second-order weak decays are the rarest processes observed to date, with measured half-lives exceeding $10^{21}$ years~\cite{Barabash2020}. Hence, detecting them demands monitoring tons of otherwise-stable atomic nuclei over several years.

$2\nu\beta\beta$-decay and $2\nu$ECEC half-lives depend on a nuclear matrix element, $M^{2\nu}$, which encodes information on the structure of the involved nuclei~\cite{Engel2017}.
Therefore, $2\nu\beta\beta$-decay measurements can test theoretical predictions of different nuclear many-body calculations of $M^{2\nu}$.
This is valuable because the same methods can also predict the nuclear matrix elements of the neutrinoless double-beta ($0\nu\beta\beta$) decay, which has not been observed yet.
Detecting $0\nu\beta\beta$ decay promises to unveil the nature of neutrinos, will establish the violation of lepton-number conservation in the laboratory and in general indicate new physics beyond the Standard Model (BSM) of particle physics~\cite{Agostini:2022zub}. Reliable $0\nu\beta\beta$ nuclear matrix elements are thus needed to extract BSM physics from measurements. However, various many-body methods currently predict matrix elements that differ by up to a factor of a few~\cite{Agostini:2022zub} and almost all calculations miss an important recently acknowledged two-nucleon operator~\cite{Cirigliano2018,Cirigliano2019,Cirigliano:2020dmx}.
Since both $2\nu\beta\beta$ and $0\nu\beta\beta$ decays share initial and final states, are sensitive to spin and isospin operators, and
may have correlated matrix elements~\cite{Horoi22,Jokiniemi22}, tests of $M^{2\nu}$ could help to reduce the theoretical uncertainties of $0\nu\beta\beta$ nuclear matrix elements.

However, calculating $M^{2\nu}$ is challenging. In fact, seldom has nuclear theory been able to predict $2\nu\beta\beta$ half-lives before their measurement; calculations yield shorter half-lives than experiment and are
corrected with an {\it ad hoc} reduction of $M^{2\nu}$, usually known as ``quenching''~\cite{Ejiri:2019ezh,Caurier12,Barea2015}. A similar correction is needed in Gamow-Teller (GT) $\beta$ decays~\cite{Wildenthal:1983zz,Chou:1993zz,MartinezPinedo96,Pirinen2015}, except in {\it ab initio} calculations which include two-body currents and many-body correlations~\cite{Gysbers2019}. Unfortunately, the latter are not yet capable of reproducing ${2\nu\beta\beta}$ decays,
even for the lightest emitter $^{48}$Ca~\cite{Novario:2020dmr,Belley:2020ejd}.
For other methods, if the deficiency is systematic~\cite{MartinezPinedo96,Chou:1993zz,Wildenthal:1983zz}, the quenching needed
for an unmeasured $2\nu\beta\beta$ decay can be inferred from other GT and $2\nu\beta\beta$ decays in the same mass region~\cite{Caurier12}. Another approach is to construct
an effective field theory (EFT) for $\beta$ decays with low-energy couplings (LECs) adjusted to GT transitions, so that $M^{2\nu}$ is predicted with theoretical uncertainties~\cite{CoelloPerez:2017xsq}. These strategies led to the predictions of the $2\nu\beta\beta$ half-life of $^{48}$Ca by the nuclear shell model~\cite{Caurier90}, and of the $2\nu$ECEC rate for $^{124}$Xe by the shell model~\cite{CoelloPerez:2018ghg}, quasiparticle random-phase approximation (QRPA)~\cite{Suhonen:2013rca,Pirinen2015} and EFT~\cite{CoelloPerez:2018ghg}, in good agreement with subsequent measurements~\cite{Balysh:1996vr,XENON:2019dti}.

$2\nu\beta\beta$ decays to excited states have already been measured in $^{100}$Mo~\cite{Barabash2020,CUPID-Mo:2022cel} and $^{150}$Nd~\cite{Barabash2020,Polischuk:2021bwl}---both without a prior
theoretical prediction---and are currently being explored 
in $^{76}$Ge~\cite{GERDA:2015naf,MAJORANA:2020shy}, $^{82}$Se~\cite{NEMO-3:2020ssy}, $^{130}$Te~\cite{CUORICINO:2011tpf}, and $^{136}$Xe~\cite{KamLAND-Zen:2015tnh,EXO-200:2015koy}. In this work, we predict the half-life of the $2\nu\beta\beta$ decay of $^{136}$Xe to the first $0^+_2$ excited state in $^{136}$Ba by several different many-body methods, following similar strategies as for $2\nu\beta\beta$ decays to the $0^+_{\text{gs}}$ ground state.
We provide the first shell-model predictions for this decay using the same Hamiltonians as in
previous $^{136}$Xe studies~\cite{Caurier12,Neacsu:2014bia,Horoi:2015tkc,KamLAND-Zen:2019imh,Jokiniemi2021}. 
We also present EFT results with systematic theoretical uncertainties for decays to the $0^+_2$ state following Refs.~\cite{CoelloPerez:2018ghg,Brase:2021uny}, thereby also extending the EFT calculations to $^{136}$Xe not included earlier. In addition, we improve previous QRPA results by using larger bases, consistent with $0\nu\beta\beta$-decay work~\cite{Jokiniemi2021}, and we update the interacting-boson-model (IBM-2) prediction by using refined model parameters ~\cite{Kotila:2016pib}.

\begin{figure}
    \centering
    \includegraphics[clip=,width=\linewidth]{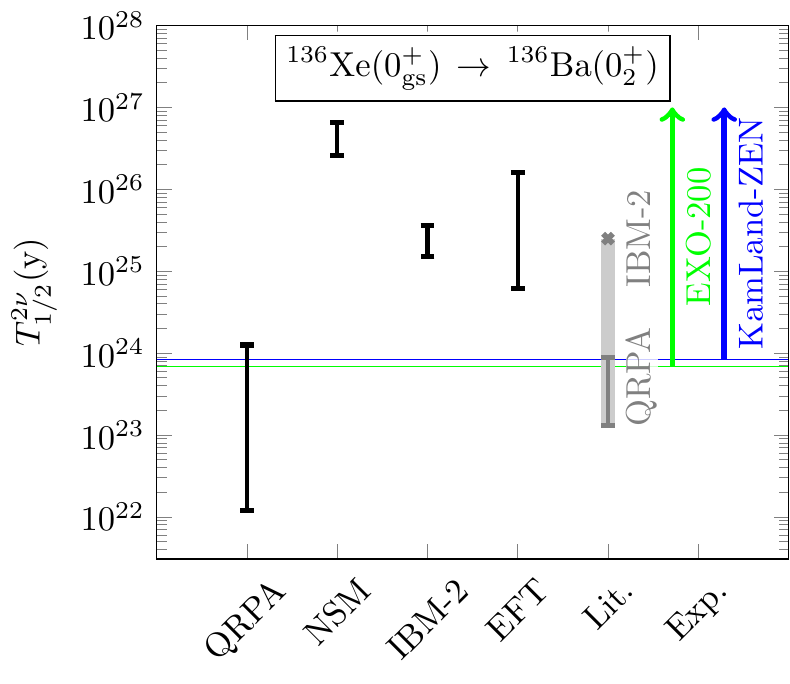}
    \caption{$^{136}$Xe $2\nu\beta\beta$-decay half-life to the $^{136}$Ba $0^+_2$ state obtained in this work (black bars) with the QRPA, nuclear shell model (NSM), IBM-2, and EFT. The results are compared with literature values (Lit.)~\cite{Pirinen2015,Barea2015} (in gray) and experimental limits~\cite{KamLAND-Zen:2015tnh,EXO-200:2015koy} (horizontal lines with arrows).}
    \label{fig:half-lives_comb_lit}
\end{figure}

Figure~\ref{fig:half-lives_comb_lit} summarizes our predictions, compared to previous works~\cite{Pirinen2015,Barea2015} and current experimental limits~\cite{KamLAND-Zen:2015tnh,EXO-200:2015koy}. While our results are consistent with experiment, only a small part of the QRPA band lies inside the non-excluded region.
In contrast, the EFT and IBM-2 favor half-lives almost an order of magnitude longer than the current experimental lower limits.
Finally, the shell model suggests that detecting the decay requires improving the current sensitivities by over two orders of magnitude. These diverse predictions indicate that the $^{136}$Xe $2\nu\beta\beta$ decay to the $0^+_2$ state in $^{136}$Ba will be
a very useful test of many-body methods used to calculate $0\nu\beta\beta$-decay nuclear matrix elements.

\section{Nuclear matrix element}

The $2\nu\beta\beta$-decay and $2\nu$ECEC nuclear matrix element is given by~\cite{Engel2017,Pirinen2015}
\begin{equation}
\label{eq:M2nu}
\begin{split}
    M^{2\nu}=&\sum_{k}\frac{\langle0^+_{ f}||\sum_a\tau^-_a\boldsymbol{\sigma}_a||1^+_k\rangle\langle1^+_k ||\sum_b\tau^-_b\boldsymbol{\sigma}_b||0^+_{i}\rangle}{(E_k-(E_i+E_f)/2)/m_e}\,,
    \end{split}
\end{equation}
where indices $a,b$ run over all nucleons, the isospin  operator $\tau^-$ turns neutrons into protons, $\boldsymbol{\sigma}$ is the spin operator, and the denominator involves the electron mass $m_e$ and the energies $E$ of the initial ($i$), final ($f$) and each $k$th intermediate $1^+_k$ state~\cite{massexcessenergies,ensdf}. The corresponding half-life
\begin{align}
\big(T^{2\nu}_{1/2}\big)^{-1}=G^{2\nu} g_A^4 \, (M^{2\nu})^2,
\label{eq:half-life}
\end{align}
also depends on a well-known phase-space factor $G^{2\nu}$~\cite{Kotila2012,Kotila:2013gea} and the axial nucleon coupling $g_{\rm A}$. For $^{136}$Xe, the phase space disfavors decays to the $0^+_2$ with respect to the $0^+_\text{gs}$ by a factor $\sim4000$.

From a measured half-life one can obtain the so-called effective matrix element~\cite{Barabash2020}:
\begin{equation}
M^{2\nu}_{\rm{eff}}=g_{\rm A}^2\,M^{2\nu}\,,
\label{eq:effective_NME}
\end{equation}
which carries all the information to be extracted from nuclear theory.
However, our shell-model, QRPA and IBM-2 results require quenching.
In these cases, we include a quenching factor $q$ obtained matching the calculated $M^{2\nu}$ to the experimentally recommended $M^{2\nu}_{\rm{eff}}$ value~\cite{Barabash2020} for ground-state-to-ground-state decay, so that we effectively use $M^{2\nu}_{\rm{eff}}=q^2g_{\rm A}^2\,M^{2\nu}$.

\section{Many-body methods}

\subsection{Quasiparticle random-phase approximation}

The QRPA method considers the $0^+$ ground states of the initial and final $\beta\beta$ nuclei as QRPA vacua, $\ket{\rm QRPA}$, building nuclear excitations on top of them. The $\beta\beta$-decay nuclear matrix elements are computed summing over intermediate states in the odd-odd nucleus [see Eq.~\eqref{eq:M2nu}], which are obtained by performing proton-neutron (pn) QRPA diagonalizations based on the initial and final states as
\begin{align}
    \ket{J_k^{\pi}}=
    \sum_{pn}\left(X^{J_k^{\pi}}_{pn}[a^{\dag}_pa^{\dag}_n]_{J}-Y^{J_k^{\pi}}_{pn}[a^{\dag}_pa^{\dag}_n]^{\dag}_{J}\right)\ket{\rm QRPA},
\end{align}
where $J_k^{\pi}$ denotes the spin-parity of the $k$th intermediate state with omitted projection quantum number $M$, $a^{\dag}$ ($\widetilde{a}$) are nucleon creation (annihilation) operators, and $X$ and $Y$ the pnQRPA backward and forward amplitudes.

On the other hand, the first excited $0^+_2$ state in $^{136}$Ba can be described in the charge-conserving (cc) QRPA formalism as a two-phonon excitation:
\begin{equation}
\ket{0^+_{\rm 2-phonon}}=\frac{1}{\sqrt{2}}[Q^{\dag}(2^+_1)Q^{\dag}(2^+_1)]_0\ket{\rm QRPA},
 \label{eq:QRPA_two-phonon}   
\end{equation}
where
\begin{equation}
\begin{split}
    \ket{2^+_1}&=Q^{\dag}(2^+_1)\ket{\rm QRPA}\,,\\
    Q^{\dag}(2^+_1)&=\sum_{a\leq b}\mathcal{N}_{ab}(X^{2^+_1}_{ab}[a^{\dag}_a \widetilde{a}_b]_2-Y^{2^+_1}_{ab}[\widetilde{a}_a a^{\dag}_b]_2)\,,
    \label{eq:QRPA_2+_1}
        \end{split}
\end{equation}
with normalization factor $\mathcal{N}_{ab}=1/\sqrt{1+\delta_{ab}}$. We follow Ref.~\cite{Hyvarinen2016} to obtain the transition densities needed to evaluate the matrix element in Eq.~\eqref{eq:M2nu} using the multiple-commutator model. This formalism includes a correction to the two-phonon transition density that was omitted in an earlier QRPA study~\cite{Pirinen2015}.

The pnQRPA is adjusted in the usual way: the particle-hole parameter $g_{\rm ph}$ reproduces the energy of the GT giant resonance in $^{136}$Cs~\cite{Jokiniemi2018}, and the particle-particle parameter $g_{\rm pp}$ is fitted to
the $2\nu\beta\beta$-decay half-life to the $0^+_{\text{gs}}$ of $^{136}$Ba, following the partial isospin-restoration scheme \cite{Simkovic2013}. We assume a quenching range $q=(0.47-1.00)$, covering typical values of the effective axial coupling $g_{\rm A}^{\rm eff}=qg_{\rm A}=(0.6-1.27)$~\cite{Pirinen2015}.
For the decay to an excited state there are two additional ccQRPA parameters. 
Due to its nature, the $2^+_1$ state is mainly sensitive to the particle-hole parameter $G_{\rm ph}$, which we adjust so that the calculated $2^+_1$ energy equals half of the measured energy of the $0^+_2,2^+_2,4^+_1$ two-phonon triplet. Since the three energies are not exactly degenerate, we obtain a range of values for $G_{\rm ph}$. For the particle-particle parameter we keep the default value $G_{\rm pp}=1.0$~\cite{Pirinen2015,Hyvarinen2016} because it does not affect much the $2^+_1$ state.

We consider large no-core single-particle bases consisting of 26 single-particle orbitals~\cite{Jokiniemi2018,Jokiniemi2021}, using a Woods-Saxon and also an adjusted basis. The latter has an increased spin-orbit splitting of the neutron $0h$ orbitals by 1.5~MeV to better reproduce excitation energies in the neighboring odd-mass nuclei.

\subsection{Nuclear shell model}

The nuclear shell model is a reference method to describe nuclear structure~\cite{MPinedo,Brown01,Otsuka19}, based on the exact diagonalization of nuclear Hamiltonians. While the configuration space is relatively small, all nuclear correlations within the space are fully captured.
The shell model has been used extensively to study GT~\cite{Chou:1993zz,Wildenthal:1983zz,MartinezPinedo96,Coraggio2020}, $2\nu\beta\beta$ and $0\nu\beta\beta$ decays~\cite{Caurier08,Caurier12,Horoi:2013jx,Horoi:2015tkc,Neacsu:2014bia,Iwata2016,Menendez2018,Coraggio2020,Jokiniemi2021,Coraggio2022,KamLAND-Zen:2019imh}. In order to describe the $2\nu\beta\beta$ decay of $^{136}\rm Xe$ we use the configuration space that comprises the single-particle orbitals $1d_{5/2}$, $0g_{7/2}$, $2s_{1/2}$, $1d_{3/2}$, and $0h_{11/2}$ for both neutrons and protons, with the GCN5082~\cite{Caurier:2010az} and QX~\cite{QiQX} interactions.
We compute the nuclear states and the $2\nu\beta\beta$ nuclear matrix elements with the code ANTOINE~\cite{FNowacki, MPinedo}. 

The low-lying energy spectra of $^{136}$Xe obtained with the GCN5082 interaction agrees very well with experiment~\cite{Vietze:2014vsa}, and for QX energies are a little too high [by $\sim200$~keV for the $2_1^+$ state, $\sim300$~keV for the $4_1^+$ and $6_1^+$ states] but also of good quality.
GCN5082 also gives a better overall description
for $^{136}$Ba, but the excitation energy of the $0^+_2$ state agrees with data in a similar way for both interactions: 1.44~MeV (GCN5082) and 1.80~MeV (QX) compared to the experimental 1.58~MeV.

For the $M^{2\nu}$ calculations we use Eq.~\eqref{eq:M2nu} through the explicit computation of $1^+$ states in $^{136}$Cs using the strength function method~\cite{MPinedo}. As it is well known, the nuclear shell model overpredicts $M^{2\nu}$ for the decay to the $0^+_{\text{gs}}$, and quenching is needed to reproduce experiment: $q=0.42$ (GCN2850)~\cite{Caurier12} and $q=0.68$~\cite{KamLAND-Zen:2019imh,Neacsu:2014bia} (QX). 
The same quenching factors are used for the decay to the $0^+_{2}$ state.

\subsection{Microscopic interacting boson model}
\label{sec:IBM}

The IBM-2 \cite{ARIMA1977205,iac87} maps the fermion Hamiltonian onto a boson space \cite{OTSUKA19781} and evaluates it with realistic bosonic wave functions. The method is discussed in detail in previous $\beta\beta$-decay studies~\cite{Barea:2009zza,Barea2015}, which include a calculation of the $^{136}$Xe $2\nu\beta\beta$ decay to the $0^+_2$ state in $^{136}$Ba~\cite{Barea2015}.
Here we improve the IBM-2 calculation using reassessed single-particle and -hole energies and interaction strengths which lead to single-particle occupancies in better agreement with nucleon-removal experiments~\cite{Kotila:2016pib}. We fit the IBM-2 parameters to reproduce the spectroscopic data of the low-lying energy states for $^{136}$Xe, and for $^{136}$Ba we take the parameters from Ref.~\cite{PUDDU1980109}. The IBM-2 configuration space is the same as for the shell model: $1d_{5/2}$, $0g_{7/2}$, $2s_{1/2}$, $1d_{3/2}$, and $0h_{11/2}$ for both neutrons and protons. One should note that the IBM-2 matrix elements to the $0^+_2$ state are sensitive to the so-called IBM-2 Majorana parameters: In Ref. \cite{PhysRevC.104.L061302}  it was found that the corresponding $^{150}$Nd $0\nu\beta\beta$-decay matrix element doubled when adjusting them to new data on $^{150}$Sm scissors- mode transitions to excited $0^+$ and $2^+$ states. Unfortunately, similar measurements for $^{136}$Ba are not available. 

The IBM-2 calculations assume the closure approximation, replacing the energies of the intermediate states in Eq.~\eqref{eq:M2nu} with an average energy $\langle E_k\rangle$, and then summing analytically over intermediate states. 
Thus, Eq.~\eqref{eq:M2nu} is simplified to 
\begin{equation}
    M^{2\nu}_{\rm IBM}=\frac{\langle0^+_{f}||\sum_{a,b}\tau^-_a \tau^-_{b}\boldsymbol{\sigma}_a\cdot\boldsymbol{\sigma}_{b}||0^+_{i}\rangle}{\left(\langle E_k\rangle-(E_i+E_f)/2\right)/m_e}\,,
    \label{eq:M^2v_GT_IBM2}
\end{equation}
where $\langle E_k\rangle=$ 11.32~MeV for $^{136}$Cs. Like in the nuclear shell model, the $2\nu\beta\beta$-decay matrix element obtained using Eq.~\eqref{eq:M^2v_GT_IBM2} is overestimated, and a quenching factor $q=0.31$ is used to agree with the empirical value \cite{Barabash2020}. We assume the same $q$ for the decay to the $0^+_\text{2}$ state.
In addition, for this decay we estimate the sensitivity to parameter changes and  model assumptions, including quenching and the closure approximation, to be $\pm 21\%$, as discussed in detail in Ref.~\cite{PhysRevC.87.014315}. 

\subsection{Effective field theory}
\label{sec:EFT}

The EFT is formulated in terms of nucleon and phonon degrees of freedom coupled to a spherical even-even core. Once the LECs are fitted to experiment (e.g., to $\beta$ decays), the EFT predicts processes dictated by the same operator (e.g., $2\nu\beta\beta$ decay). In addition to nuclear properties, the EFT systematically provides the associated theoretical uncertainties based on its power counting~\cite{EFTqunc}. 
The EFT describes well GT transitions to excited states and $2\nu\beta\beta$ decays~\cite{CoelloPerez:2017xsq}, including the prediction of the $2\nu$ECEC half-life of $^{124}$Xe~\cite{CoelloPerez:2018ghg}. Recently it has been applied to $0\nu\beta\beta$ decay as well~\cite{Brase:2021uny}.

\begin{figure}
    \centering
    \includegraphics[clip=,width=\linewidth]{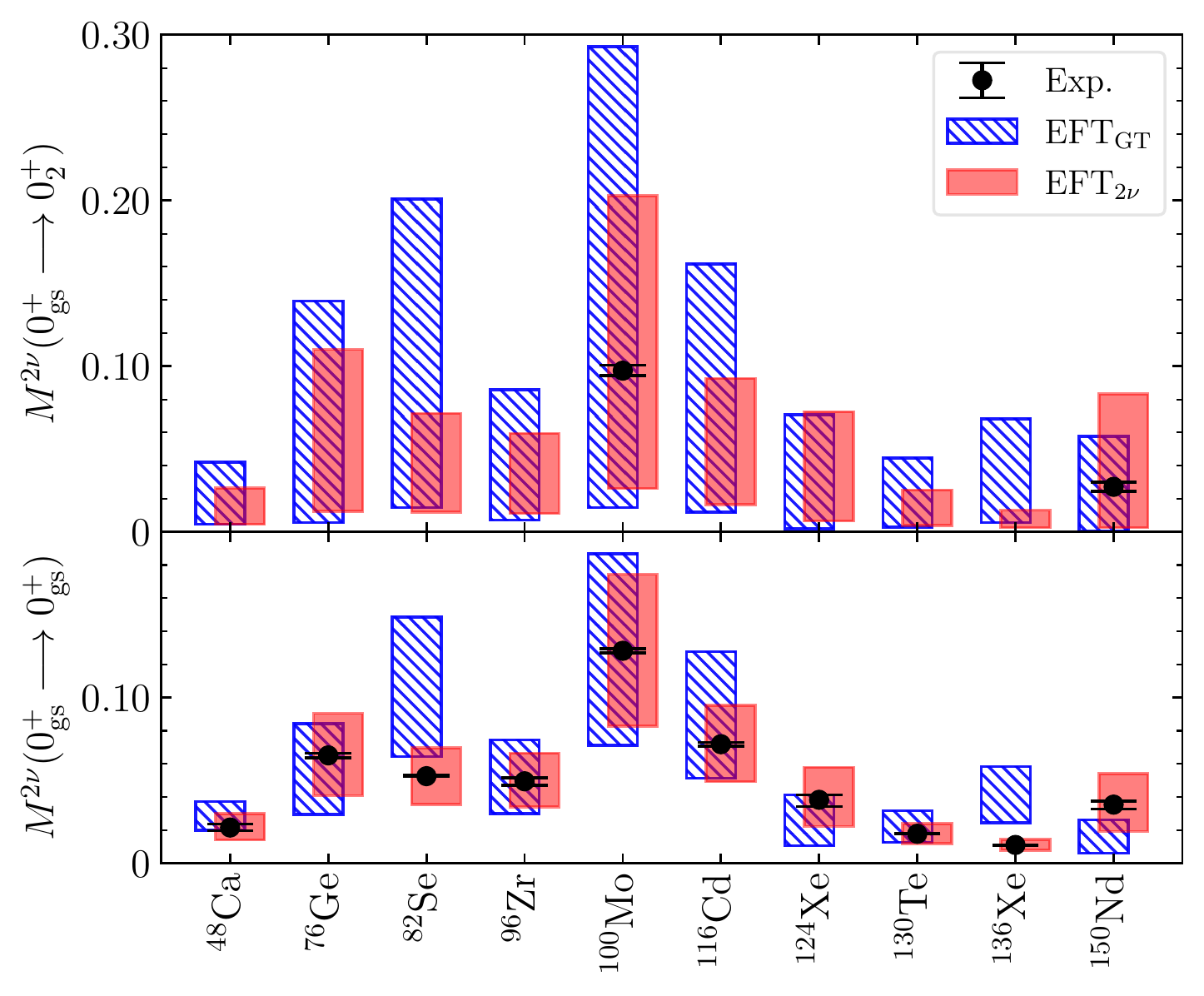}
    \caption{EFT $2\nu\beta\beta$ nuclear matrix elements ($2\nu$ECEC for $^{124}$Xe) with theoretical uncertainties at leading order, obtained with LECs fitted to GT data as in Ref.~\cite{CoelloPerez:2018ghg} (hatched blue bars)~\cite{GTstrength48Ca,GTstrength76Ge,GTstrength82Se,GTstrength96Zr,GTstrength96Mo,GTstrength130Te,GTstrength136Xe,GTstrength150Nd} and to $2\nu\beta\beta$ half-lives to the $0^+_{\text{gs}}$, the strategy of this work (solid red bars)~\cite{Barabash2020,XENON:2019dti}. The nuclear matrix elements are compared to empirical values (black circles)~\cite{Barabash2020,XENON:2019dti}. Bottom: $0^+_{\text{gs}}$ to $0^+_{\text{gs}}$ decays. Top: $0^+_{\text{gs}}$ to $0^+_2$ decays.}
    \label{fig:res}
\end{figure}

In the EFT, $M^{2\nu}$ can be calculated using the single-state-dominance approximation~\cite{CoelloPerez:2017xsq}. Previous EFT studies fitted the LECs to $\beta$ decay or GT strengths~\cite{CoelloPerez:2017xsq,CoelloPerez:2018ghg}, which for $^{136}$Xe are limited to the GT strength from the $^{136}$Xe$(^3$He,t)$^{136}$Cs charge-exchange reaction~\cite{GTstrength136Xe}. We label this matching EFT$_{\rm GT}$. The bottom panel of Fig.~\ref{fig:res} (blue bands) shows that, even though this strategy works very well for most of the measured decays within uncertainties~\cite{CoelloPerez:2017xsq,CoelloPerez:2018ghg}, it overpredicts the $^{136}$Xe $2\nu\beta\beta$ nuclear matrix element to the $0^+_{\text{gs}}$.

We attribute this deficiency to the fact that the low-energy spectrum of $^{136}$Xe does not exhibit the properties of a spherical collective system~\cite{massexcessenergies}. In contrast, $^{136}$Ba shows a one-phonon excited state at approximately half of the energy of the two-phonon excitations ($\sim$ 818/1565 keV) and typical ps lifetimes for the collective excitations~\cite{massexcessenergies}. However this nucleus cannot be used to fit the LECs due to lack of GT data. For 
$^{48}$Ca and $^{96}$Zr the EFT also better describes the final nuclei, but in these cases there is GT information available involving $^{48}$Ti and $^{96}$Mo to fit the LECs. Figure~\ref{fig:res} shows that for these decays the EFT agrees well with experiment, but the agreement is lost (for $^{48}$Ca) or worsens (for $^{96}$Zr), if the LECs are only fitted to GT data from the initial nuclei, as in $^{136}$Xe.
Finally, the EFT does not describe well the $^{150}$Nd $2\nu\beta\beta$ decay to the $0^+_{\text{gs}}$, probably due to its deformation.

Therefore, we here follow an alternative strategy, labelled EFT$_{2\nu}$, in contrast to the previous EFT$_{\text{GT}}$. In the EFT with single-state-dominance approximation the matrix element for the decays into the $0^+_2$ state and $0^+_{\text{gs}}$ can be related by \cite{CoelloPerez:2017xsq}
\begin{equation}\label{eq:2nGTgs02}
    \begin{aligned}
        &M_{\mathrm{EFT}}^{2\nu}(0^+_\mathrm{gs}\rightarrow 0^+_2)\approx\\
        &\left(1+\frac{D_{10^+_2}}{D_{20^+_2}}+\frac{D_{10^+_2}}{D_{30^+_2}}\right) \frac{D_{10^+_\mathrm{gs}}}{D_{10^+_2}}\frac{\sqrt{2}}{3} \, M_{\mathrm{EFT}}^{2\nu}(0^+_\mathrm{gs}\rightarrow 0^+_\mathrm{gs})\,,
    \end{aligned}
\end{equation}
where $D_{kf}=E_k-(E_i+E_f)/2$. The analytical expression of the associated uncertainty is given by Eq.~(44) in Ref.~\cite{CoelloPerez:2017xsq}. 
We then fit the LECs directly to measured $2\nu\beta\beta$ decays effectively through its $M_{\mathrm{EFT}}^{2\nu}$ matrix element, assigning an EFT uncertainty that includes the single-state-dominance approximation and the uncertainties in $M_{\mathrm{EFT}}^{2\nu}$ from the EFT truncation and the experimental values. By construction, the EFT$_{2\nu}$ reproduces the empirical data for the decay to the $0^+_{\text{gs}}$, see bottom panel in Fig.~\ref{fig:res} (red bands).

Then, we use Eq.~\eqref{eq:2nGTgs02} to predict the decays to the $0^+_{2}$ excited state in the EFT. The top panel in Fig.~\ref{fig:res} shows that this approach agrees well with the two measured decays in $^{100}$Mo and $^{150}$Nd. Therefore, we use the EFT$_{2\nu}$ with its leading order uncertainties as recommended values in this work for $^{136}$Xe. Note that for the decay to the $0^+_{2}$ state the EFT$_{2\nu}$ and the EFT$_{\text{GT}}$ are consistent within uncertainties.

\section{Results and discussion}

Table~\ref{tab:t12_136Xeexc} gives the predicted half-life of the $^{136}{\rm Xe}$ $2\nu\beta\beta$-decay to the $0^+_2$ state of $^{136}{\rm Ba}$ obtained with the QRPA, nuclear shell model, IBM-2, and EFT. Figure~\ref{fig:half-lives_comb_lit} shows the combined  ranges for each many-body method. All of them have been adjusted, either via LECs or a quenching factor, to reproduce the known $2\nu\beta\beta$ decay to the $0^+_{\rm gs}$. For the QRPA, the uncertainty range for each basis is dominated by the $G_{\text{ph}}$ and quenching ranges considered. In the shell model, the band for each Hamiltonian is simply given by the empirical $M^{2\nu}_\text{eff}$~\cite{Barabash2020} used to obtain $q$.
The estimated IBM-2 and calculated EFT$_{2\nu}$ uncertainties are explained in Secs.~\ref{sec:IBM} and~\ref{sec:EFT}, respectively.

\setlength{\tabcolsep}{8.5pt}
\begin{table}[t]
\caption{$^{136}{\rm Xe}$ $2\nu\beta\beta$-decay half-life to the $0^+_2$ state of $^{136}{\rm Ba}$, in units of the scale of the decay to the ground state, $T^{2\nu}_{1/2}=2.2\cdot10^{21}$~y~\cite{Barabash2020}. We give ranges for the QRPA with Woods-Saxon (WS) and adjusted (adj.) bases, the nuclear shell model (NSM) with the GCN5082 and QX interactions, our IBM-2 results, the EFT with its leading-order uncertainties, as well as previous QRPA and IBM-2 results.}
\label{tab:t12_136Xeexc}
    \centering
    \begin{tabular}{ccc}
    \toprule
         Method & Reference & $T^{2\nu}_{1/2}$ ($10^{21}$~y) \\
         \midrule
         QRPA (adj.) & This work & $(0.14-2.9) \cdot10^2$          \\
         QRPA (WS) & This work &$(0.47-13)\cdot10^2$          \\
         QRPA & \cite{Pirinen2015} & $(1.3-8.9) \cdot10^2$ \\ \midrule
         NSM (GCN)  & This work & $(2.5-2.9)\cdot10^5$\\
         NSM (QX) & This work & $(5.8-6.6)\cdot10^5$\\ \midrule   
         IBM-2 & This work & $(1.5-3.6)\cdot 10^{4}$\\
         IBM-2 & \cite{Barea2015} & $2.5 \cdot10^4$\\ \midrule
         EFT$_{2\nu}$& This work & $(0.62-16)\cdot 10^{4}$\\ 
         \bottomrule
    \end{tabular}
\end{table}

The dispersion of the theoretical predictions is striking:
except for the IBM-2 and EFT$_{2\nu}$---which 
are fully consistent within uncertainties---different many-body methods predict inconsistent half lives ranging
more than four orders of magnitude. Thus, in the extremes, the shell-model and QRPA matrix elements can disagree by up to a factor 100. This difference is much more pronounced than the one for $0\nu\beta\beta$-decay matrix elements \cite{Agostini:2022zub}, which does not exceed a factor five or so.

\begin{figure}[t]
\centering
\includegraphics[clip=,width=\linewidth]{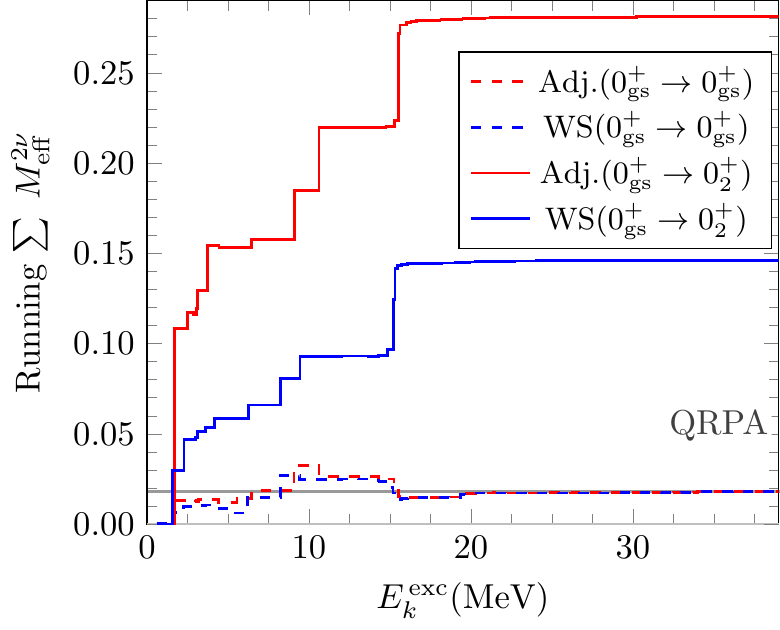}
\caption{$M^{2\nu}_{\rm eff}$ running sum as a function of the excitation energy of the intermediate state, $E_k^\text{\,exc}$, for the $^{136}$Xe $2\nu\beta\beta$ decay to the $0^+_\text{gs}$ (dashed lines) and $0^+_2$ states (solid lines) in $^{136}$Ba. QRPA results obtained with Woods-Saxon (WS, blue) and adjusted (adj., red) bases and $g_{\rm A}^{\rm eff}=1.0$ ($q=0.79$). The horizontal gray band shows the empirical value~\cite{Barabash2020}.}
\label{fig:NMEs_gs_QRPA}
\end{figure}

\begin{figure}[t]
\centering
\includegraphics[clip=,width=\linewidth]{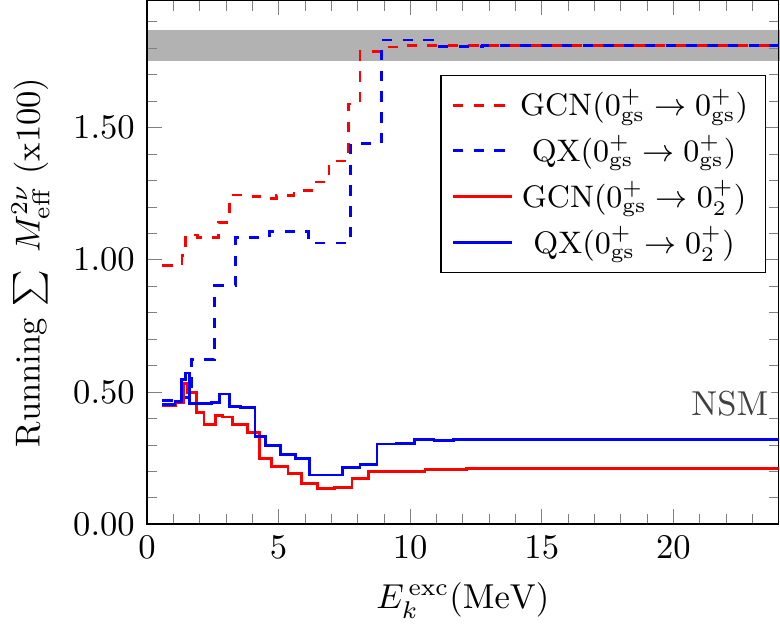}
\caption{Same as Fig.~\ref{fig:NMEs_gs_QRPA} (y-axis scale multiplied by 100) comparing shell-model results obtained with the GCN5028~\cite{Caurier:2010az} (red) and QX~\cite{QiQX} (blue) interactions.}
\label{fig:NMEs_1exc_NSM}
\end{figure}

In order to understand these results, Figs.~\ref{fig:NMEs_gs_QRPA} and~\ref{fig:NMEs_1exc_NSM} show the running sum of the effective matrix element $M^{2\nu}_{\text{eff}}$ as a function of the excitation energy of the intermediate $1^+$ states in $^{136}$Cs for the QRPA and shell model, respectively. References~\cite{KamLAND-Zen:2019imh,Simkovic2018} performed a similar analysis limited to the decay to the $0^+_{\rm gs}$. Figure~\ref{fig:NMEs_gs_QRPA} shows that for the QRPA, for the decay to the $0^+_2$ state all intermediate states contribute with the same sign, contrary to the decay to the $0^+_{\text{gs}}$ where there are some cancellations at high energies $E\sim10$~MeV, as previously pointed out in Ref.~\cite{Simkovic2018}.
In addition, the individual contributions to the decay to the excited state are also larger, especially in the case of the adjusted basis; here the contribution of $1^+$ states around 2~MeV exceeds the summed contribution of all intermediate states for the ground-state decay. Therefore, the QRPA matrix element is much larger for the decay to the $0^+_2$ state, leading to a relatively short half-life. This is especially the case for the adjusted basis, whose prediction is in strong tension with experimental limits, see Table~\ref{tab:t12_136Xeexc} and Fig.~\ref{fig:half-lives_comb_lit}.

In contrast, Fig.~\ref{fig:NMEs_1exc_NSM} shows that for the shell model there are no cancellations between contributions to $M^{2\nu}_{\text{eff}}$ for the decay to the $0^+_{\text{gs}}$, while intermediate states add up with different signs in the decay to the $0^+_2$ state. The qualitative behaviour of the two running sums is similar for the two Hamiltonians.
As a consequence, the shell-model matrix element is significantly smaller for the decay to the $0^+_2$ state, and the corresponding half-life becomes very long. Note that while Fig.~\ref{fig:NMEs_gs_QRPA} shows that QRPA matrix elements receive contributions up to intermediate states with energies $E\sim15$~MeV, the shell-model matrix elements in Fig.~\ref{fig:NMEs_1exc_NSM} are converged at $E\sim10$~MeV. This suggests that the shell model could be missing contributions from high-energy states due to its limited configuration space. If these potential contributions had positive sign, like in the QRPA, they would shorten the half-life of the decay to the $0^+_2$ state predicted by the shell model.

On the other hand, the EFT with single-state dominance does not allow for strong cancellations in the matrix elements, since possible cancellations are just part of the theoretical uncertainty.
This is in contrast with our analysis of the QRPA (decay to the $0^+_{\text{gs}}$) or shell-model (decay to the $0^+_2$ state) running sums in Figs.~\ref{fig:NMEs_gs_QRPA} and~\ref{fig:NMEs_1exc_NSM}. 
This qualitative difference in the running sums cannot be assessed in the lower-resolution EFT using the single-state dominance approximation. For instance, high-energy $1^+$ states which are relevant for the cancellations of the QRPA and NSM running sums may not be fully captured by the EFT. Nonetheless, the agreement with measured decays to $0^+_2$ states shown in Fig.~\ref{fig:res} suggests that the EFT theoretical uncertainties at least in part capture the uncertainty associated with this approximation.
Due to absence of explicit cancellations, the EFT decay to the $0^+_{\text{gs}}$ and $0^+_2$ states can be expected to be relatively similar. The IBM-2 matrix elements for the two decay branches obtained in the closure approximation are also similar. This naturally leads to the EFT and IBM-2 intermediate half-lives in Table~\ref{tab:t12_136Xeexc} and Fig.~\ref{fig:half-lives_comb_lit}, about $10^3-10^4$ times longer than the decay to the $0^+_\text{gs}$ as dictated by the different phase-space factors.

\section{Summary}

In summary, we have predicted the half-life of the $^{136}{\rm Xe}$ $2\nu\beta\beta$ decay to the $0^+_2$ state of $^{136}{\rm Ba}$ using four different many-body methods that are also used to calculate $0\nu\beta\beta$-decay nuclear matrix elements. Our results indicate a large uncertainty from nuclear theory: while the QRPA prediction is close to current limits, the nuclear shell model indicates a half-life more than two orders of magnitude longer. The IBM-2 and EFT results lie in between. We have provided error estimates for all four methods, but only the EFT ones can be considered as systematic theoretical uncertainties. Our findings thus highlight that further experimental searches of this decay are very useful tests of theoretical models aiming to predict $0\nu\beta\beta$ nuclear matrix elements. Finally, for the EFT we have also presented results for $2\nu\beta\beta$ decays to the excited $0^+_2$ state in other nuclei,  which agree well within uncertainties for the two measured cases.

\section*{Acknowledgements}

This work was supported by the Arthur B.~McDonald Canadian Astroparticle Physics Research Institute, the Academy of Finland (Grant Nos.~314733 and 345869), by the European Research Council (ERC) under the European Union's Horizon 2020 research and innovation programme (Grant Agreement No.~101020842 and No.~951281), by the ``Ram\'on y Cajal'' program with grant RYC-2017-22781, and grants CEX2019-000918-M, PID2020-118758GB-I00 and and RTI2018-095979-B-C41 funded by MCIN/AEI/10.13039/501100011033 and by ``ESF Investing in your future''.
TRIUMF receives funding via a contribution through the National Research Council of Canada.

\bibliography{biblio}

\end{document}